\documentclass[preprint,12pt]{elsarticle}

\usepackage{amssymb}
\usepackage{subcaption}
\usepackage{caption}
\usepackage{xcolor}
\usepackage{soul}
\usepackage{multicol}
\usepackage{multirow}
\usepackage{graphicx}
\usepackage{ulem}

\graphicspath{ {./} }

\journal{ArXiv}

\begin{document}

\begin{frontmatter}



%\title{Multiple sclerosis lesion segmentation framework for brain MRI}
\title{Boosting multiple sclerosis lesion segmentation through attention mechanism}

\author[inst1]{Alessia Rondinella}
\author[inst2]{Elena Crispino}
\author[inst1]{Francesco Guarnera}
\author[inst1]{Oliver Giudice}
\author[inst1]{Alessandro Ortis}
\author[inst3]{Giulia Russo}
\author[inst5]{Clara Di Lorenzo}
\author[inst4]{Davide Maimone}
\author[inst3]{Francesco Pappalardo}
\author[inst1]{Sebastiano Battiato}

\affiliation[inst1]{organization={Department of Mathematics and Computer Science, University of Catania},%Department and Organization
            addressline={Viale Andrea Doria 6}, 
            city={Catania},
            postcode={95125}, 
            state={Italy}}
\affiliation[inst2]{organization={Department of Biomedical and Biotechnological Sciences, University of Catania},%Department and Organization
            addressline={Via Santa Sofia 97}, 
            city={Catania},
            postcode={95125}, 
            state={Italy}}
            
\affiliation[inst3]{organization={Department of Drug and Health Sciences, University of Catania},%Department and Organization
            addressline={Viale Andrea Doria 6}, 
            city={Catania},
            postcode={95125}, 
            state={Italy}}
            
\affiliation[inst4]{organization={Centro Sclerosi Multipla, UOC Neurologia, ARNAS Garibaldi},%Department and Organization
            addressline={P.zza S. Maria di Gesù}, 
            city={Catania},
            postcode={95124}, 
            state={Italy}}
            
\affiliation[inst5]{organization={UOC Radiologia, ARNAS Garibaldi},%Department and Organization
            addressline={P.zza S. Maria di Gesù}, 
            city={Catania},
            postcode={95124}, 
            state={Italy}}
            
\begin{abstract}

Magnetic resonance imaging is a fundamental tool to reach a diagnosis of multiple sclerosis and monitoring its progression. Although several attempts have been made to segment multiple sclerosis lesions using artificial intelligence, fully automated analysis is not yet available. State-of-the-art methods rely on slight variations in segmentation architectures (e.g. U-Net, etc.). However, recent research has demonstrated how exploiting temporal-aware features and attention mechanisms can provide a significant boost to traditional architectures.  
This paper proposes a framework that exploits an augmented U-Net architecture with a convolutional long short-term memory layer and attention mechanism which is able to segment and quantify multiple sclerosis lesions detected in magnetic resonance images. Quantitative and qualitative evaluation on challenging examples demonstrated how the method outperforms previous state-of-the-art approaches, reporting an overall Dice score of 89\% and also demonstrating robustness and generalization ability on never seen new test samples of a new dedicated under construction dataset.

\end{abstract}




\end{frontmatter}


\section{Introduction}
\label{sec:introduction}
\begin{figure}[t]
        \centering
        \begin{subfigure}[b]{0.23\textwidth}
                \centering
                \includegraphics[width=\textwidth]{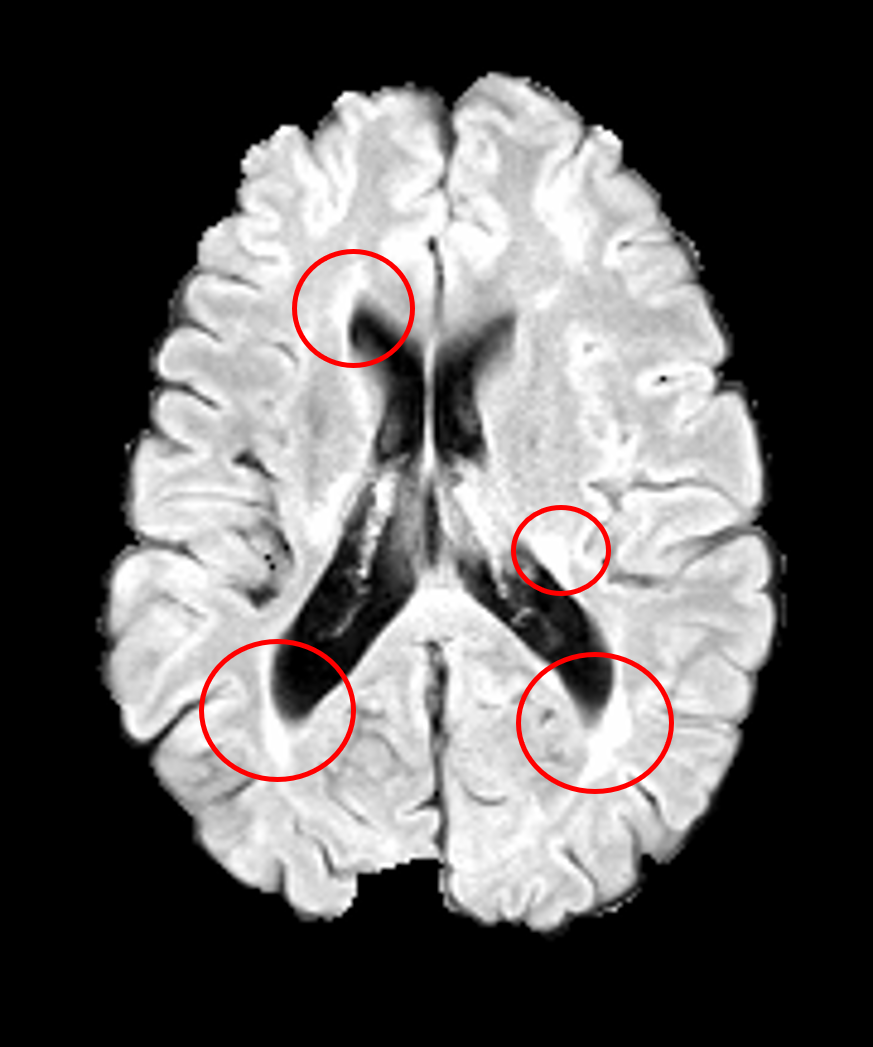}
                \caption{}
                \label{fig:a}
        \end{subfigure}
        \begin{subfigure}[b]{0.23\textwidth}
                \centering
                \includegraphics[width=\textwidth]{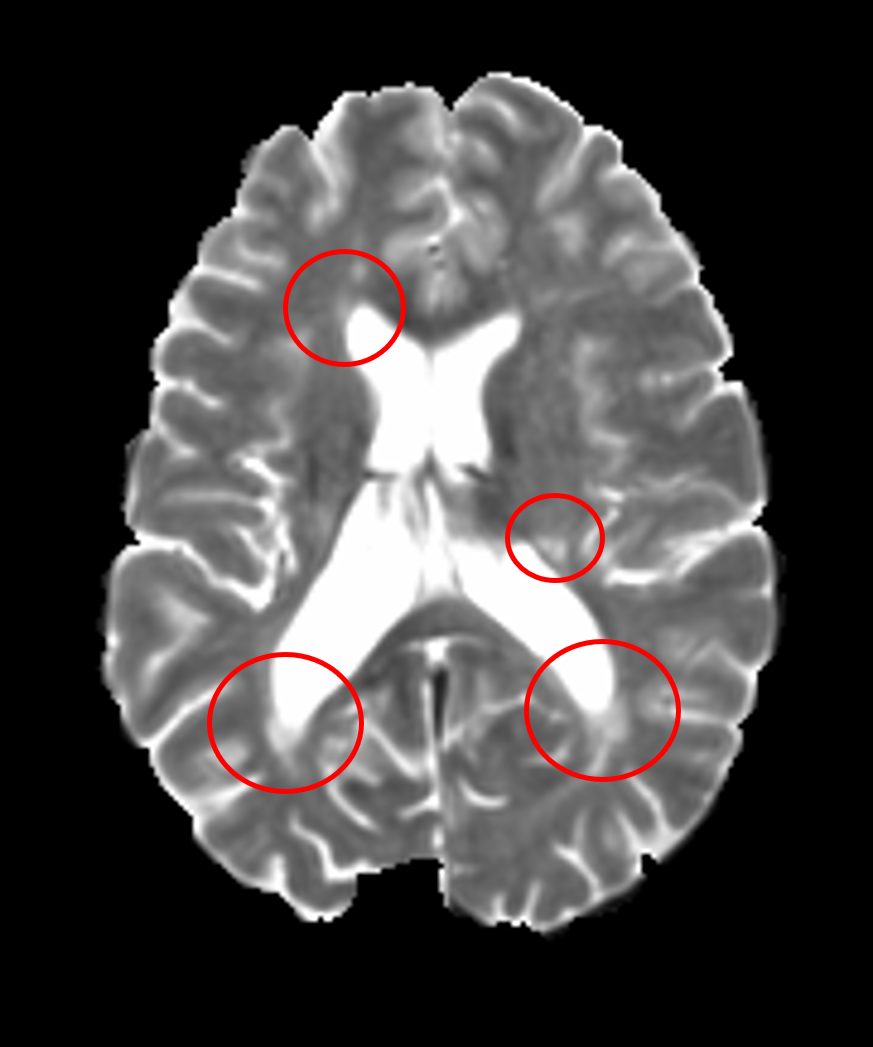}
                \caption{}
                \label{fig:b}
        \end{subfigure}
        \begin{subfigure}[b]{0.23\textwidth}
                \centering
                \includegraphics[width=\textwidth]{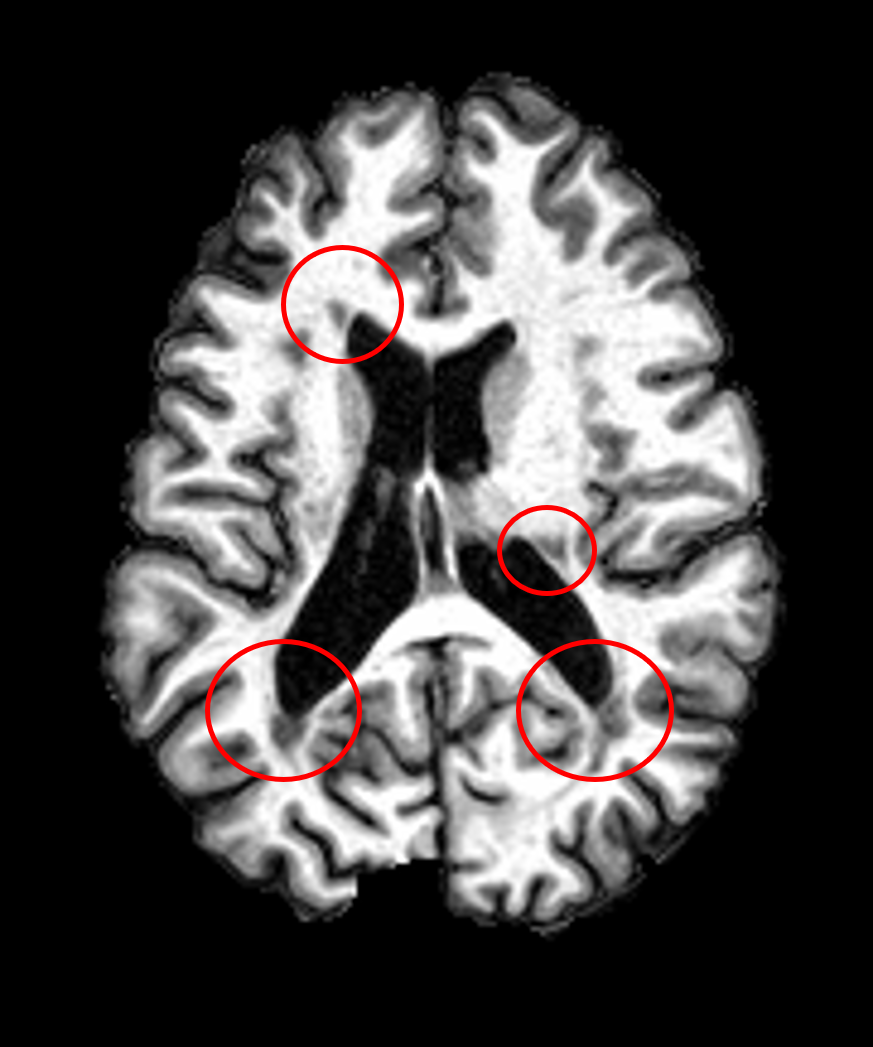}
                \caption{}
                \label{fig:c}
        \end{subfigure}
        \begin{subfigure}[b]{0.23\textwidth}
                \centering
                \includegraphics[width=\textwidth]{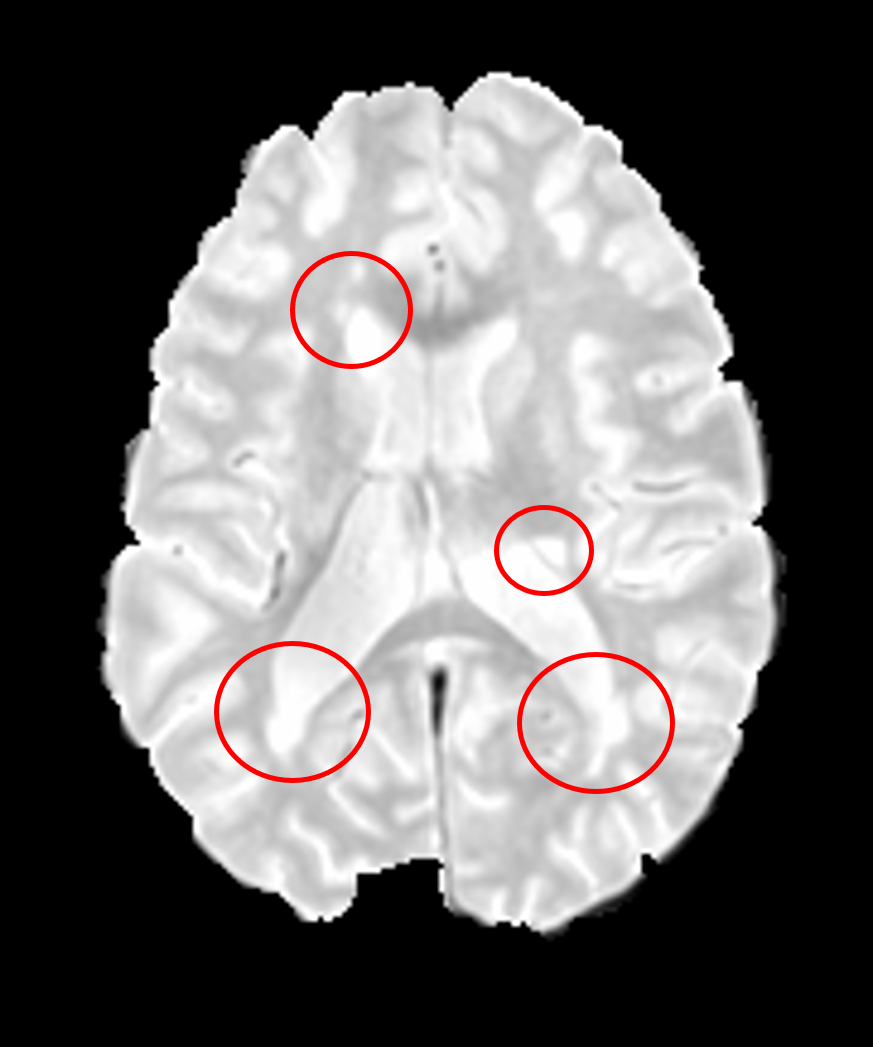}
                \caption{}
                \label{fig:d}
        \end{subfigure}
        \caption{Sample axial MRI images of the brain of an MS patient in each modality of acquisition showing MS lesion in (a) Flair, (b) T2-weighted, (c) T1-weighted and (d) PD-weighted. Red circles highlight lesions in the different modality of acquisition.} \label{fig:scans2}
\end{figure}

Multiple Sclerosis~(MS) is a chronic inflammatory demyelinating disease  of the Central Nervous System~(CNS) \cite{lassmann2018multiple}, with neuropathologic features characterized by focal areas of inflammation with myelin and axonal loss.
 MS lesions may be detected in vivo by Magnetic Resonance Imaging~(MRI) in different areas of the brain and the spinal cord and they accumulate over time \cite{filippi2018}. %\cite{steinman1996multiple}. 
 Selective localization of lesions on MRI (periventricular, cortical/iuxtacortical, brain stem/cerebellar, and spinal cord) is also relevant for the diagnosis of MS and detection of new or enlarging lesions at follow-up, and is routinely used in the evaluation of therapeutic response and disease progression \cite{thompson2018diagnosis, wattjes20212021}.
Manual annotation of MS lesions on MRI scans is a time consuming task and requires substantial efforts by specialized experts. Moreover, inter and intra operator variability is unavoidable and may affect accuracy and reproducibility of lesion segmentation \cite{molyneux1999visual}. 
Thus, there is an increasing interest today in automation of MRI reading and evaluation to avoid the bias introduced by human raters and to make this information available for routine clinical practice \cite{shoeibi2021applications}.
Typically, the longitudinal brain MRI protocol involves distinct kinds of sequences, which generate different types of images that vary according to the contrast of the various tissues that compose the brain. The most common MRI sequences used to detect MS lesions are the Fluid Attenuated Inversion Recovery~(FLAIR), T1-weighted, T2-weighted, and PD-weighted images. In the T1- weighted sequence, white matter appears lighter than gray matter, and cerebrospinal fluid~(CSF) appears dark. In the T2-weighted sequence, the white matter appears darker than the gray matter, while the CSF appears bright. FLAIRs images are like T2s, except that CSF is suppressed.
MS lesions appears hypointense in T1-w and hyperintense in T2-w, PD-w and FLAIR sequences, with respect to normal tissue intensities. Lesions are most detectable in the FLAIR images, where they appear hyperintense and usually well distinguishable from surrounding tissues. Figure \ref{fig:scans2} shows four MRI brain images for each different acquisition types with MS lesions~(MS lesions are pointed by red circle). In our method, similarly to other works in the field \cite{khayati2008novel, khayati2008fully, gessert2020multiple}, the most discriminating MRI sequence (FLAIR) was exploited.

 The starting point of our study is one of the most widely used networks in the state of the art for this purpose, the U-Net architecture \cite{ronneberger2015u}, which is widely used not only in medical image segmentation, but also in general segmentation tasks. In this work, an extended Fully Convolutional DenseNet~(FC-DenseNet) \cite{jegou2017one} for MS lesion segmentation is proposed; it follows the U-Net structure \cite{ronneberger2015u} with the addiction of Long Short-Term Memory~(LSTM) layer and extensive usage of attention mechanisms to detect FLAIR-w MS lesions in longitudinal brain MRI.
Attention \cite{vaswani2017attention} is a technique that aims to mimic the cognitive attention of humans by enforcing neural networks to pay greater attention to most informative input data and ignore the rest.
Attention mechanisms have been shown to be effective in capturing global dependencies and have become an integral part of semantic segmentation tasks \cite{sinha2020multi}. The FC-DenseNet has been properly extended with an attention mechanism based on the usage of squeeze and attention blocks \cite{zhong2020squeeze}, in order to accentuate the group of pixel from the same classes employing different spatial scales. Squeeze and Attention blocks~(SA) represent a component that can be easily incorporated within the backbone, able to improve network performance through operations  applied on both local and global level. Moreover, the space propagation of the lesions with a similar shape between adjacent images suggested to introduce a Long short-term memory (LSTM) layer \cite{shi2015convolutional}; it permits to preserve spatial information between longitudinal axis of data.

The performance of the proposed architecture was evaluated employing a cross-validation scheme in patients with lesions on follow-up scans.
%The proposed model combines features of the U-Net architecture, which is widely used in medical image analysis, including contraction and expansion stages, with the integration of attention mechanisms that enhance their representational ability and guide the network's focus to salient regions.
The architecture described in Figure \ref{fig:model} represents the best results of some tested models described in the ablation studies (Section \ref{sec:ablation}).

The training phase of a Deep Neural Network architecture typically requires a large amount of labeled images \cite{lecun2015deep}. A relevant issue in MRI lesions segmentation is the presence of just few small example in each dataset available \cite{shoeibi2021applications} and the lack of homogeneity between different repositories, due to the usage of different scanners and/or acquisition protocols. This makes the segmentation challenging, raising concerns about the results obtained from the different methods, which are difficult to compare and generalize to other datasets. For these reasons, we are actually working on the generation of a new labeled MRI dataset as part of the \textit {``In Silico World (ISW): Lowering barriers to ubiquitous adoption of In Silico Trials``} (Grant agreement ID: 101016503, PROGRAMME: H2020-EU.3.1. - SOCIETAL CHALLENGES - Health, demographic change and well-being, CALL: H2020-SC1-DTH-2018-2020). It will be larger than the actual ones, with heterogeneous samples (patients with different stages of disease) and with labeled MS lesions validated by employing different experts. 
The proposed method, its future extensions and the under construction labeled dataset will be included into the Universal Immune System Simulator~(UISS). UISS is a multi-compartment,  multi-scale, polyclonal, stochastic, and patient-specific agent-based model~(ABM) that is able to simulate immune system dynamics both in physiological and pathological scenarios \cite{sips2022silico}. 
UISS simulator framework has been extended to model MS pathogenesis and its interaction with the host immune system [18], taking into account both cellular and molecular entities. Particularly, UISS- MS takes into account B cells, T helper (CD4+ T cells), T cytotoxic (CD8+ T cells), conventional dendritic cells (DCs), macrophages (M), plasma B cells (P cells), immunocomplexes (IC), oligodendrocytes (ODC), interferon-gamma (IFN-G), interleukins of type x (IL-x), transforming growth factor beta (TGFB), myelin basic proteins (MBP), immunoglobulins class G (IgG) and chemokines (as generic chemokines) [19]. For each modeled patient, the age at MS onset, baseline MRI lesion load, oligoclonal bands status, and the administered treatment are usually considered.
%\sout{By default, the three main compartments represented are the thymus, the bone marrow, and a portion of a generic secondary organ. A specific volume of peripheral lymph nodes or blood is represented by a lattice, and in each lattice point many entities, namely “agents”, may be present. UISS takes into account both cellular and molecular entities; cellular entities are often tracked individually and modeled as solitary agents, while molecular entities are considered by their concentration inside the lattice point. UISS simulator framework has been extended to model MS pathogenesis and its interaction with the host immune system \cite{pappalardo2018agent}. Particularly, UISS-MS takes into account B cells, T helper (CD4+ T cells), T cytotoxic (CD8+ T cells), conventional dendritic cells (DCs), macrophages (M), plasma B cells (P cells), immunocomplexes (IC), oligodendrocytes (ODC), interferon-gamma (IFN-G), interleukins of type x (IL-x), transforming growth factor beta (TGFB), myelin basic proteins (MBP), immunoglobulins class G (IgG) and chemokines (as generic chemokines) \cite{pappalardo2020potential}. For each modeled patient, the age at MS onset, baseline MRI lesion load, oligoclonal bands status, and the administered treatment are usually obtained.}

A limit of the current UISS-MS framework is that only qualitative data about MRI lesion load have been inserted \cite{russo2021computational}. For this reason, the quantitative data about the MRI lesion load obtained with the framework here proposed will be integrated into the UISS framework, with the aim to represent and predict the disease progression of MS patients as well as to more realistically simulate the immune response to specific treatments.

%%The remainder of this paper is organized as in the following. In Section~\ref{sec:stateoftheart} the state-of-the-art is presented; Section~\ref{sec:methodology} explains the dataset employed and the proposed method. Experimental results are reported in Section~\ref{sec:experiments}, while Section~\ref{sec:conclusion} concludes the paper.

The remainder of this paper is organized as follows. Section~\ref{sec:stateoftheart} resumes the state-of-the-art of MS lesions segmentation while Section~\ref{sec:methodology} explains the employed dataset and the proposed method. Experimental results with state-of-the-art comparisons and ablation studies are reported in Section~\ref{sec:experiments}, whereas Section~\ref{sec:conclusion} concludes the paper.

\section{State of the art}
\label{sec:stateoftheart}
In recent years, various Artificial Intelligence (AI) methods based on deep learning have been proposed for the classification, detection, and segmentation of health-related conditions from medical images. For example, \cite{alhatemitransfer} utilizes deep learning methods for the classification of stroke in MR images, whereas \cite{savacs2022comparison} compares the classification performance of several deep learning architectures in ultrasound images for early diagnosis of carotid artery disease.
Moreover, deep learning-based architectures have been utilized for the segmentation of various organs and tissues in medical images, including autoimmune disease segmentation using histopathological images \cite{singh2023data}, lung segmentation for Covid-19 prediction \cite{wang2020prior, napoli2022mixup, rondinella2023attention} or automatic segmentation of MS lesions from MRI scans;
 however, the results obtained are still far from those generated by manual segmentation.
Furthermore, semi-automatic or automatic approaches have proven to be sensitive to MRI variability and different acquisition modalities, leading to a loss of accuracy. The results to date are still distant from those of human experts despite the enormous efforts.
Recently, MS lesion segmentation methods have been classified into main categories, most of which include unsupervised, supervised and deep learning-based methods.
Regarding supervised approaches,the authors of \cite{roura2015toolbox} proposed a method where lesions were segmented by applying an intensity threshold to the FLAIR image. In \cite{knight2016ms} a combination of a fuzzy classification method with an edge-based method is used, and the segmentation was obtained applying thresholding and a false-positive reduction technique.
Regarding unsupervised approaches, \cite{schmidt2012automated} proposed an algorithm for MS lesion detection based on the intensity distribution of the three different tissues to detect lesions. In \cite{freifeld2007lesion}, the authors used a probabilistic model, Gaussian Mixture Model~(GMM) to delineate lesion contours. The work in \cite{khayati2008fully} proposed a framework by exploiting a Bayesian classifier and Markov Random Field~(MRF) model to compute the a-priori probability for each tissue class.
Most of the latest works exploit deep neural network-based methods. In particular, most of the published works employ methods based on Convolutional Neural Network~(CNN) and U-Net architectures. In \cite{schmidt2019automated}, the authors proposed an automated pipeline for serial analysis of MS lesions using FLAIR scans, relying on cross sectional segmentation of lesions in white matter. In \cite{gabr2020brain}, a multiclass FCNN model is proposed for brain tissue segmentation (gray matter, white matter, and cerebrospinal fluid) and MS lesions in T2-W scans. A framework for FLAIR segmentation is proposed in \cite{placidi2021multiple} by training two CNNs on MSSEG-2016 dataset in the axial, coronal, and sagittal directions.
In \cite{raab2022multimodal} the authors use a multimodal 2D U-Net, encoding the different image modalities in separate downsampling channels, while \cite{kamraoui2022deeplesionbrain} propose a combination of 3D networks for a spatially distributed strategy robust to domain-shifting.

The employing of attention has shown interesting improvements in some fields of medical image segmentation. In a recent work \cite{hashemi2022delve}, segmentation of MRI FLAIR and T2 images is performed using a modified U-Net and Attention U-Net, proposing the fusion of the masks obtained from a better segmentation of Flair and T2.
Another study \cite{sarica2023dense} proposed a new dense residual U-Net model that leverages attention gate and channel attention techniques to improve the performance of automated MS lesion segmentation in MRI, while the authors in \cite{gessert2020multiple} propose a CNN based on two-paths architecture with the addition of a attention-driven interaction block between them able to share information between two different time points. 
Recent works demonstrated how the right use of attention in MS domain could significantly improve the results. The authors in \cite{shoeibi2021applications} summarize recent researches on automated MS diagnosis based on Deep Learning~(DL) and AI analyzing the features exploited, the preprocessing techniques employed and the challenges faced by published works, in part exploited in the current proposal.    

%%utilizing MRI modalities and deep learning approaches.
%It was these recent works that motivated the present work regarding the use of attention in the MS domain, which, starting from \cite{hashemi2022delve}, will find a better way to improve the performance of the state of the art in MS lesion segmentation in Flair scans using attention.\color{black}
%\color{blue}AGGIUNGI PARAGRAFO O CAPITOLO DEDICATO ALL'ATTENTION ED ALL'IMPORTANZA DELLA STESSA NEI TASK DI SEGMENTATION \color{red}MEGLIO CITARE LAVORI CHE USANO ATTENTION NELLA SEGMENTAZIONE SEMANTICA? \color{black}

\section{Methodology}
\label{sec:methodology}
\subsection{Image dataset}
\label{sec:dataset}
 
\begin{figure}[t]
\centering
\begin{subfigure}{.3\textwidth}
  \centering
  % include first image
  \includegraphics[width=\linewidth]{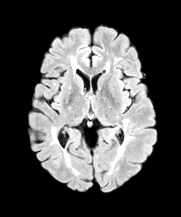}  
  \caption{}
  \label{fig:slice}
\end{subfigure}
\begin{subfigure}{.3\textwidth}
  \centering
  % include second image
  \includegraphics[width=\linewidth]{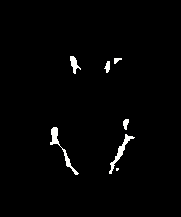}  
  \caption{}
  \label{fig:mask}
\end{subfigure}
\caption{An example of FLAIR image from the ISBI-2015 dataset (a) with the corresponding mask annotated by rater 1 (b).}
\label{fig:slice-groundtruth}
\end{figure}

The dataset employed in our method is a subset of the ISBI2015 challenge dataset; it is a public available set of images presented at the Longitudinal MS Lesion Segmentation Challenge \cite{carass2017longitudinal}, organized in conjunction with the ISBI 2015 conference. 
The full dataset is composed of $19$ patients MRI scans, acquired at multiple time points on a 3.0 Tesla MR scanner, but only 5 patients are available with the corresponding segmentation mask. Each patient presents two different segmentation masks, produced by two expert human raters; it is important to note how in many cases the masks are different, which explain the difficult of the task also in presence of MS expert. The $14$ patients without segmentation masks were originally used to validate the challenge algorithm but were discarded in our case for obvious reasons. An example of image belonging to the dataset and its relative mask is shown in Figure \ref{fig:slice-groundtruth}.

The selected $5$ patients were acquired at different time points: $4$ of them have $4$ time points longitudinal scans, where the last has $5$ time points, for a total of $21$ different acquisitions; the time interval between two consecutive acquisitions is approximately $1$ year. To note that the course of multiple sclerosis is highly variable and follow-up scans do not necessarily correspond to disease progression, as MS lesions may appear at different times and in different parts of the brain. Each acquisition contains the original MR images, the images after co-registration (geometric alignment of the images), brain extraction and non-uniformity correction, and the masks representing the MS lesion. Each scan contains different images sequences: T1-weighted, T2-weighted, PD-weighted, and FLAIR. 
To assess the stability of the model, we performed our experiments by evaluating our method only on the masks labeled by rater 1. 
To perform our experiments, only the FLAIR images were employed because MS lesions in white matter appear hyperintense and are more visible than other types of sequences; every FLAIR sequence is composed of $181$ images of size $181 \times 217$.

%\color{blue}
%stiamo testando il metodo su altri dataset e che stiamo creando un dataset molto più numeroso che pubblicheremo a breve e su cui testeremo questo metodo e anche un metodo 3D data la numerosità delle immagini e dove non ci sarà la necessità delle immagini baseline come input.
%\color{black}

We are currently generating a dataset containing MRI scans, acquired as part of project \textit {``In Silico World (ISW): Lowering barriers to ubiquitous adoption of In Silico Trials``}, with MS lesions labeled by two experts. This dataset will contain scans of numerous patients acquired at multiple time points; MR images were acquired by a 1.5 T scanner (Ingenia, Philips MR Systems, Release 4.1.3.2,
Best, The Netherlands) under a regular maintenance program and sequences employed to reveal MS lesions were: 3D T2-FLAIR, Axial T2-FSE and 3D T1-gradient echo.
An addictional test of the proposed method with an item of the future dataset was carried-out and presented in Section \ref{sec:additionaltest}.
The study was approved by the corresponding Hospital Ethics Committee and all patients gave their informed consent.

%We will test the method presented in this paper on this new dataset. In addition, given the numerosity of the new dataset, we will test 3D methods and methods in which the baseline of the patient will not be necessary.

\subsection{Preprocessing}
\label{sec:preprocessing}
\begin{figure}[t]
\centering
\includegraphics[width=\textwidth]{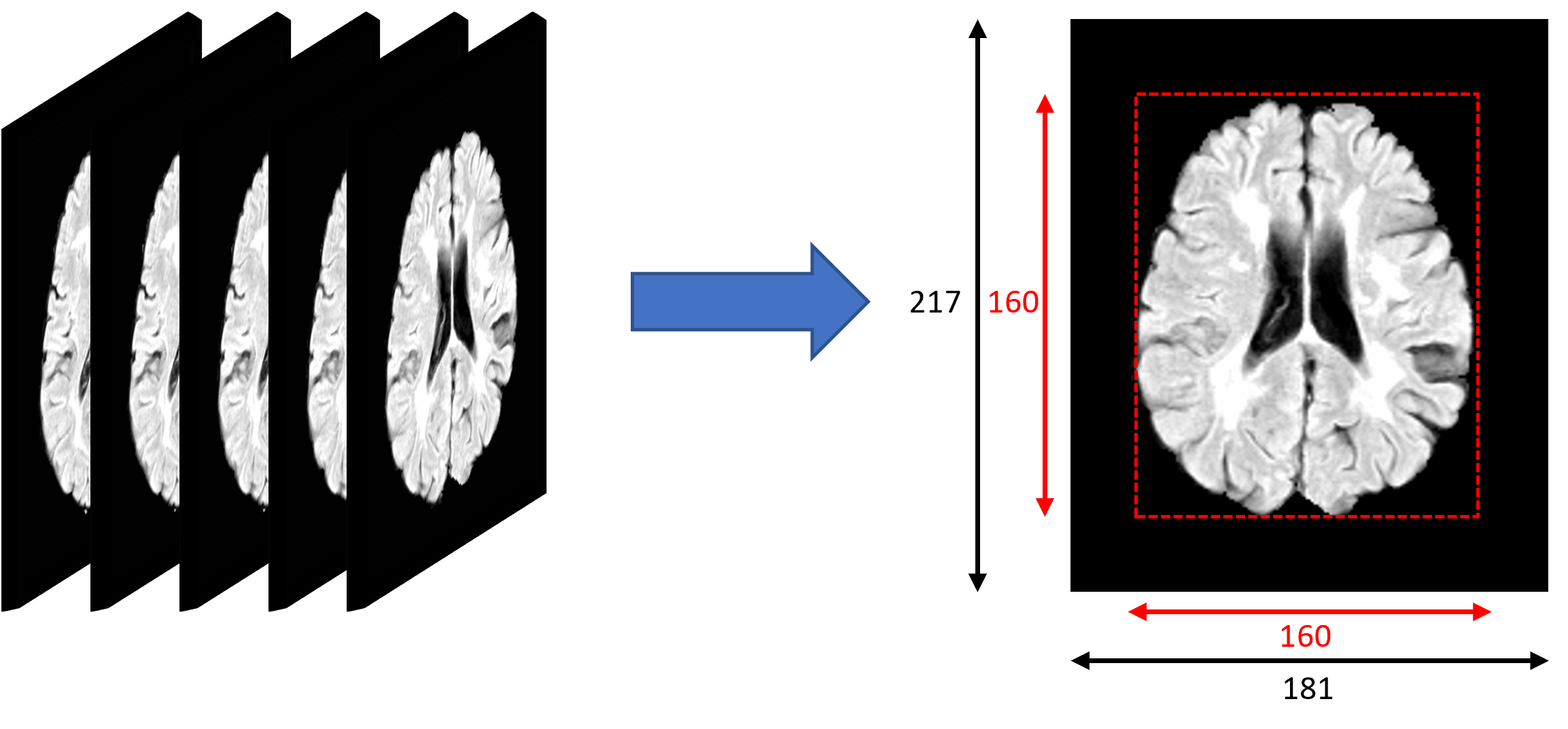}
\caption{Preprocessing performed on the FLAIR slices. First the entire black masks were removed, then the edges were cropped; the resulting image from the preprocessing steps has dimensions $160 \times 160$.}
\label{fig:preprocessing_images}
\end{figure}

State-of-the-art applied different image preprocessing methods, as co-registration \cite{collignon1995automated}, intensity correction \cite{zheng2009improvement}\cite{popescu2012optimizing}, skull-stripping \cite{roura2014marga}\cite{shattuck2001magnetic}.
To avoid processing absolutely useless or marginal information, the removal of black images on terminal parts of each scan, where lesions are not present is usually performed. For the same reason, Hashemi et al. in \cite{hashemi2022delve} applied a removal of the black part outside the brain, where is not possible to find lesion areas, also in images containing lesions, in order to give only ``active`` information to the network. Although the action could seem obvious, the removal of these areas improves the results significantly. Figure \ref{fig:preprocessing_images} shows an example of the aforementioned preprocessing: it was applied in all the considered items of our method reducing the input images at $160 \times 160$ pixels. This action is  also useful to overcome the imbalance inside the ground-truth masks (Figure \ref{fig:mask}): our goal is to identify the lesions, which are identified by white pixels, while the rest of the image is identified by black pixels (a binary classification). It is evident how in the mask image (Figure \ref{fig:mask}) the number of pixels of white area (target) is sensibly less than black ones enforcing the model to better predict dominant areas. Removing the whole black masks and resizing the image also allow, as a side effect, to significantly reduce the overall training time.

%Figure \ref{fig:preprocessing_images} shows the preprocessing performed on the images in the ISBI2015 dataset. For each patient, the same preprocessing steps was applied on both baseline and follow-up images. These steps follow the work done by \cite{hashemi2022delve} on the same dataset. Because there are images in each scan that don't have MS lesions, the first step was to remove the black masks, which do not contain lesion information.

%The next step is cut the margin areas, which is totally black in all images, and it is possible to consider it as a background.
%The reason for this type of preprocessing is to overcome the dataset imbalance: our goal is to identify the lesions, which are identified by white pixels, while the rest of the image is considered to be the background, identified by black pixels. Therefore, we have only two classes; it is a binary segmentation.
%In these segmentation masks the number of pixels of white area (targeted lesion) is less than number of black pixels. Hence, the dominant class, of black pixels, will be more easily predicted by the model. To improve minority class performance and reduce class imbalance, we remove the black pixels that are not useful for lesion segmentation purposes. In addition, eliminating entire black masks significantly reduces the training time. The network processes all images as PNG files based on the input size set by default to 224x224. 
After the above-mentioned preprocessing steps, we obtain images with corresponding square masks of size $160 \times 160$, that constitute the input to the network.

\subsection{Proposed model}
\label{sec:model}
In recent years Medical Imaging researchers demonstrated how U-Net and its customized architectures \cite{ronneberger2015u} provided effective results in various scenarios. The capacity of U-net to produce detailed segmentation maps, using a very limited amount of data, makes it particularly helpful; it assumes relevance in the context of medical imaging since access to large amounts of labeled data is very limited. 

The state-of-the-art employed U-Net in medical image segmentation adding some customization to enhance the results; in our case squeeze and attention block \cite{zhong2020squeeze}, as reported in next sections, sensibly improve the results on MS lesion segmentation.
Before introducing the overall architecture, the main blocks added to our segmentation network will be described below.

\paragraph{\textbf{Squeeze-and-attention module}}
Squeeze-and-attention~(SA) modules \cite{zhong2020squeeze} attempt to emphasize channels that contain informative features and suppress the non-informative ones. This module performs a re-weighting technique that pay attention locally and globally; locally because the convolutional operations are performed in a small pixel neighborhood, while globally they selects which image feature maps to focus on to perform segmentation. 
SA extends the feature recalibration operations performed by the squeeze-and-excitation~(SE) modules \cite{hu2018squeeze} to not apply fully-squeezed operation to spatial information.

\paragraph{\textbf{Convolutional LSTM}}
Convolutional LSTM \cite{shi2015convolutional}, combines the advantages of RNN and CNN architectures. It introduces convolutional layers in place of fully connected layers in an LSTM to enable more structure in the recurrent levels. In medical image segmentation, spatial information is essential to be able to reconstruct an entire area. For this reason, a convolutional LSTM uses the convolution operator in recurrent connections to learn the spatial features of adjacent images.

\begin{figure}[h]
\centering
\includegraphics[width=0.9\textwidth]{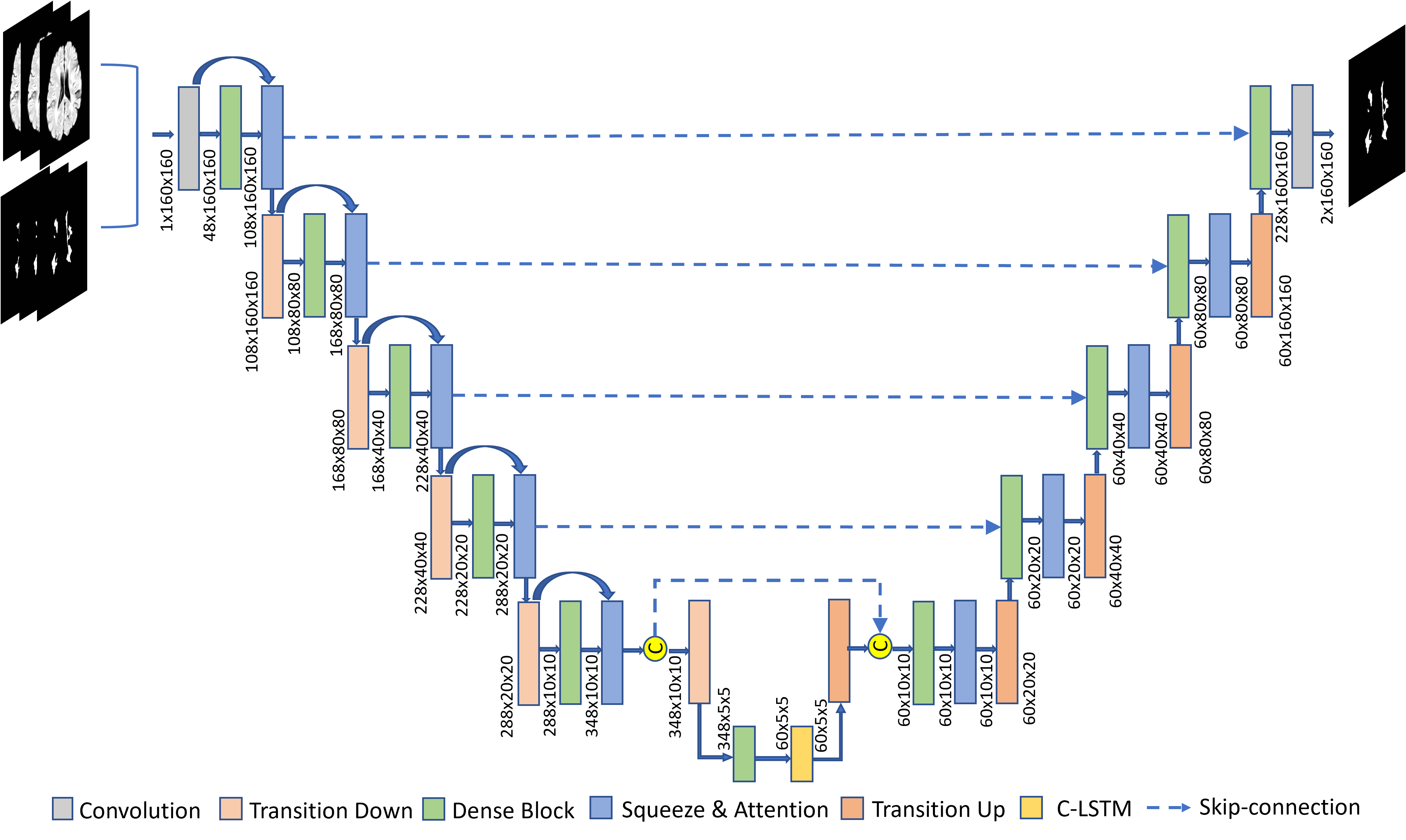}
\caption{Architecture of the FLAIR MS lesion segmentation model. This model takes as input a sequence of 3 FLAIR slices and returns as output the corresponding segmentation map.}
\label{fig:model}
\end{figure}

%\color{blue}
%BISOGNA AGGIUNGERE LA PARTE CHE SPIEGA CHE LA RETE HA BISOGNO DI ESSERE TRAINATA CONOSCENDO IL PAZIENTE. QUESTA COSA POTREBBE "INIBIRE IL REVIEWER" PER CUI DOBBIAMO SUBITO DIRE CHE "COME FATTO DA \cite{hashemi2022delve}...." E POI SPIEGARE IL PERCHE' (ABBIAMO POCHI ITEMS, SOLO 21, QUINDI LA RETE DEVE PRE-CONOSCERE IL PAZIENTE) 
%\color{black}

\hfill \break
Figure \ref{fig:model} shows the proposed MS Lesion segmentation architecture. It is build starting from a Fully Convolutional Densely Network (known as Tiramis\'u network \cite{jegou2017one}) based on a modified U-Net structure \cite{ronneberger2015u}. As mentioned before, compared to the conventional U-net architectures for MS lesion segmentation \cite{zhang2019multiple}, squeeze and attention blocks to both the downsampling and upsampling path were added to emphasize the more informative feature maps; also a unidirectional convolutional LSTM \cite{shi2015convolutional} was inserted in the bottleneck, in order to catch spatial correlation of sequentially axial slices. 
These slices are processed independently in the first part of the net (convolutional operations) and combined in the network bottleneck to produce the final output. In particular, the segmentation result of the central slice is obtained providing to the network an input with the same slice and the two adjacent ones, (previous and next): it is easy to observe how lesions are propagating over the space with a similar shape. We chose the number of 3 slices since the MS lesion is more likely to be within this spatial sequential and because in a greater sequence the lesion could lead not only to considering lesions with different structure, but also increase the training time. 

The architecture consists of a downsampling path composed of a convolutional layer and five sequences composed of: a dense block, a squeeze-attention block and transition down blocks. The upsampling path is symmetrical to the downsampling one and also each couple upsampling/downsamplig block is concatenated through Skip connections.

\begin{table*}[htbp]
\tiny
	\centering
	\resizebox{\linewidth}{!}{
	\renewcommand{\arraystretch}{1.1}
		\begin{tabular}{c||c||c||c}
		\hline

		    \hline
		    \hline
		    
		    \textbf{Position} &\textbf{Layer} &\textbf{Composed by} &\textbf{\# layers}\\
		    \hline
      Downsampling & Conv2d & - & 1\\
      %\hline
      %Downsampling & DenseBlock & DenseLayer (x5) & 5\\
       \hline
			\multirow{5}{*}{Downsampling} & \multirow{5}{*}{DenseBlock} & \textbf{DenseLayer} & \multirow{5}{*}{5}\\
   
			\cline{3-3}
			 &  & \textbf{DenseLayer} & \\
    
    \cline{3-3}
			 &  & \textbf{DenseLayer} & \\
    \cline{3-3}
			 &  & \textbf{DenseLayer} & \\
    \cline{3-3}
			 &  & \textbf{DenseLayer} & \\

     \hline
			\multirow{4}{*}{Downsampling} & \multirow{4}{*}{SqueezeAttentionBlock} & AvgPool2d & \multirow{4}{*}{5}\\
			\cline{3-3}
			 &  & \textbf{conv\_block} & \\
    \cline{3-3}
			 &  & \textbf{conv\_block} & \\
    \cline{3-3}
			 &  & Upsample & \\

      \hline
			\multirow{5}{*}{Downsampling} & \multirow{5}{*}{TransitionDown} & BatchNorm2d & \multirow{5}{*}{5}\\
			\cline{3-3}
			 &  & ReLU & \\
    \cline{3-3}
			 &  & Conv2d & \\
    \cline{3-3}
			 &  & Dropout2d & \\
    \cline{3-3}
			 &  & MaxPool2d & \\

%\hline
      %Bottleneck & DenseBlock & DenseLayer (x5) & 1\\
       \hline
			\multirow{5}{*}{Bottleneck} & \multirow{5}{*}{DenseBlock} & \textbf{DenseLayer} & \multirow{5}{*}{1}\\
			\cline{3-3}
			 &  & \textbf{DenseLayer} & \\
    \cline{3-3}
			 &  & \textbf{DenseLayer} & \\
    \cline{3-3}
			 &  & \textbf{DenseLayer} & \\
    \cline{3-3}
			 &  & \textbf{DenseLayer} & \\

\hline
      Bottleneck & ConvLSTM & Conv2d & 1 \\
\hline
      Upsampling & TransitionUp & ConvTranspose2d & 5 \\

%\hline
      %Bottleneck & DenseBlock & DenseLayer (x5) & 1\\
       \hline
			\multirow{5}{*}{Upsampling} & \multirow{5}{*}{DenseBlock} & \textbf{DenseLayer} & \multirow{5}{*}{5}\\
			\cline{3-3}
			 &  & \textbf{DenseLayer} & \\
    \cline{3-3}
			 &  & \textbf{DenseLayer} & \\
    \cline{3-3}
			 &  & \textbf{DenseLayer} & \\
    \cline{3-3}
			 &  & \textbf{DenseLayer} & \\

\hline
			\multirow{4}{*}{Upsampling} & \multirow{4}{*}{SqueezeAttentionBlock} & AvgPool2d & \multirow{4}{*}{5}\\
			\cline{3-3}
			 &  & \textbf{conv\_block} & \\
    \cline{3-3}
			 &  & \textbf{conv\_block} & \\
    \cline{3-3}
			 &  & Upsample & \\
 \hline
      Upsampling & Conv2d & - & 1\\
       \hline
  Exit & Softmax & - & 1\\

    \hline
			\multirow{4}{*}{} & \multirow{4}{*}{\textbf{DenseLayer}} & BatchNorm2d & \multirow{4}{*}{}\\
			\cline{3-3}
			 &  & ReLU & \\
    \cline{3-3}
			 &  & Conv2d & \\
    \cline{3-3}
			 &  & Dropout2d & \\
     \hline
			\multirow{6}{*}{} & \multirow{6}{*}{\textbf{conv\_block}} & Conv2d & \multirow{6}{*}{}\\
			
    \cline{3-3}
			 &  & BatchNorm2d & \\
    \cline{3-3}
			 &  & ReLU & \\
    \cline{3-3}
			 &  & Conv2d & \\
     \cline{3-3}
			 &  & BatchNorm2d & \\
    \cline{3-3}
			 &  & ReLU & \\
    \cline{3-3}

			\hline
			\hline
    \multicolumn{4}{c}{Total parameters = 13,242,782 }\\
                \hline
			\hline

		\end{tabular}
	}
	\caption{FC-DenseNet + SA + C-LSTM layers for FLAIR MS lesion segmentation: every row shows the position,  the number of times usages and how it is composed.\\
    \textbf{DenseLayer} and \textbf{conv\_block} are complex blocks, explained in the last two rows.}
	\label{table:table1}
\end{table*}

The proposed network consists of over 400 levels. As it is not possible to include a table listing all the levels, these are described using the grouping in Table \ref{table:table1}. As mentioned earlier, the network takes FLAIR images as input because lesions are more visible in this modality, generating the lesion segmentation mask as output.
Within the architecture each slice is shown in grayscale, in which every pixel value contains only information on intensity. The group of three slices were passed through a standard convolutional layer, which is needed to increase the size of feature maps. 
Then they go through the downsampling path, also called encoder, consisting of a sequence of dense blocks, squeeze-attention blocks, and transition down blocks.
In the downsampling process, the spatial resolution of the images was gradually reduced and the number feature maps were gradually increased.
The advantage of applying squeeze-attention modules is to emphasize channels that contain informative features and suppress all other non-informative ones. Specifically, it is demonstrated that the SA block introduces a pixel-group attention mechanism with a convolutional attention channel, which allows the network to selectively focus on the most significant groups of pixels in the input image, while excluding other groups. This is achieved through spatial attention, where neighboring pixels of the same class are grouped together and treated as a single unit during processing, allowing for pixel-wise prediction \cite{zhong2020squeeze}. In particular, in the case of multiple sclerosis lesion segmentation, it is important to consider the relationships between pixels in a group, as lesions often have a distinct shape and structure.
In this path, the output feature maps from each transition down level are concatenated with the output feature maps of each squeeze and attention level, and used as the input of the next level.
The downsampling path is followed by the bottleneck, which is tipically characterized by a sequence of levels that process the slices when they have the lowest possible spatial resolution. It consists of a dense block and a unidirectional convolutional LSTM layer. A 2D convolutional approach was chosen instead of 3D, as the dataset employed had a limited number of samples available. By incorporating the LSTM layer, the network can capture the spatial dependencies between adjacent slices, leading to better feature representations and improved accuracy in the final output, thereby focusing on the sequentiality of the scans instead of giving an entire scan per single step. By doing so, the sequential task has many more samples than the 3D task.
At the end of the bottleneck there is the upsampling path, also called the decoder, which is symmetrical to the downsampling path and is useful for recovering the input spatial resolution that is lost during the previous path. The spatial resolution of the images is then gradually increased and the number of feature maps is gradually reduced.
The main characteristics of the upsampling path is the presence of skip connections, which concatenate the future maps at the exit of each Transition Up blocks with those that have the same resolution coming from the downsampling path to create the input of the next layer.
The skip connection were useful to recover spatially detailed information lost during the downsampling path. At the end of the upsampling path, there is a convolutional layer and the softmax which encodes for each pixel a probability for each possible class. Thus, the output of the model is the segmentation mask of the central slice of the sequence.

\section{Experiments and results}
\label{sec:experiments}
\subsection{Evaluation metrics}
\label{sec:metric}

The evaluation of the model was done comparing the predicted segmentation masks with the reference ones that, as mentioned previously, were chosen by only one of the experts as ground truth.

%In our experiments, we evaluated our model by comparing automatically segmented lesion masks with manually delineated masks contained in the dataset. We consider only the annotation of rater 1 (mask1) as the ground truth.
%In order to demonstrate the effectiveness of the proposed method, several experiments considering different combination of the dataset were carried out.
As evaluation metrics, Dice score, sensitivity, specificity, Extra Fraction, Intersection Over Union (IOU), Positive Predictive Value (PPV) and Negative Predictive Value (NPV) were used. The Dice score \cite{dice1945measures} is defined in Eq.(\ref{eq:dsc}), 
%\begin{center}
% DSC = $\frac{2*TP}{2*TP+FP+FN}$  
%\end{center}
\begin{equation}\label{eq:dsc}
     DSC = \frac{2*TP}{2*TP+FP+FN}
\end{equation}
where TP, FP and FN denote the number of True Positive, False Positive and False Negative pixels, respectively.
Dice score is a metric used to measure the similarity between two classes, widely used in medical image segmentation.

The Sensitivity is defined in Eq.(\ref{eq:sens}), 
\begin{equation}\label{eq:sens}
  SENS = \frac{TP}{TP+FN}
\end{equation}
Sensitivity measures the number of positive voxel that are properly identified.

Specificity is defined in Eq.(\ref{eq:spec}), 
\begin{equation}\label{eq:spec}
  SPEC = \frac{TN}{TN+FP}
\end{equation}
Specificity measures the number of negative voxel that are properly identified.

Extra Fraction (EF) is defined in Eq.(\ref{eq:extra}), 
\begin{equation}\label{eq:extra}
  EF = \frac{FP}{TN+FN}
\end{equation}
Extra Fraction measures the number of voxels segmented that are not in the reference segmentation.

Intersection Over Union (IOU) is defined in Eq.(\ref{eq:iou}), 
\begin{equation}\label{eq:iou}
  IOU = \frac{TP}{TP+FN+FP}
\end{equation}
Intersection Over Union measures the number of voxels segmented that quantifies the degree of overlap between two region. 

Positive Predictive Value (PPV) is defined in Eq.(\ref{eq:ppv}), 
\begin{equation}\label{eq:ppv}
  PPV = \frac{TP}{TP+FP}
\end{equation}
Positive Predictive Value measure the number of positive voxel that are true positive results.

Negative Predictive Value (NPV) is defined in Eq.(\ref{eq:npv}), 
\begin{equation}\label{eq:npv}
  NPV = \frac{TN}{TN+FN}
\end{equation}
Negative Predictive Value measure the number of negative voxel that are true negative results.

\subsection{Experimental setup and hardware specification}
\label{sec:setup}
As explained in Section \ref{sec:methodology} the experiments were performed employing images obtained after the pre-processing phase, removing the black areas. Given the low number of patients (only $5$) and scans (only $21$), the tests were carried out through a cross-validation strategy considering 5-fold, with $17$ scans to train the network, $3$ scans used as validation set and a scan to test it. As done by Hashemi et al. in \cite{hashemi2022delve} part of patient's scans were employed during training phase while the remaining ones for testing.

\begin{itemize}
\item \textit{Fold1} includes a total of 1119 images as a training set, 197 images as a validation set and 70 images as a test set (patient 1 at T4).
%%\item \textit{Fold1} includes a total of 1243 images as a training set, 73 images as a validation set and 70 images as a test set (patient 1 at T4).
\item \textit{Fold2} includes a total of 1119 images as a training set, 183 images as a validation set and 84 images as a test set (patient 2 at T4).
%%\item \textit{Fold2} includes a total of 1218 images as a training set, 84 images as a validation set and 84 images as a test set (patient 2 at T4).
\item \textit{Fold3} includes a total of 1119 images as a training set, 200 images as a validation set and 67 images as a test set (patient 3 at T4).
%%\item \textit{Fold3} includes a total of 1252 images as a training set, 67 images as a validation set and 67 images as a test set (patient 3 at T4).
\item \textit{Fold4} includes a total of 1136 images as a training set, 208 images as a validation set and 42 images as a test set (patient 4 at T4).
%%\item \textit{Fold4} includes a total of 1306 images as a training set, 38 images as a validation set and 42 images as a test set (patient 4 at T4).
\item \textit{Fold5} includes a total of 1111 images as a training set, 225 images as a validation set and 50 images as a test set (patient 5 at T4).
%%\item \textit{Fold5} includes a total of 1285 images as a training set, 51 images as a validation set and 50 images as a test set (patient 5 at T4).
\end{itemize}

\begin{figure}[htbp]
\centering
        %%Fold1
        \begin{subfigure}{0.41\textwidth}
                \centering
                \includegraphics[width=\textwidth]{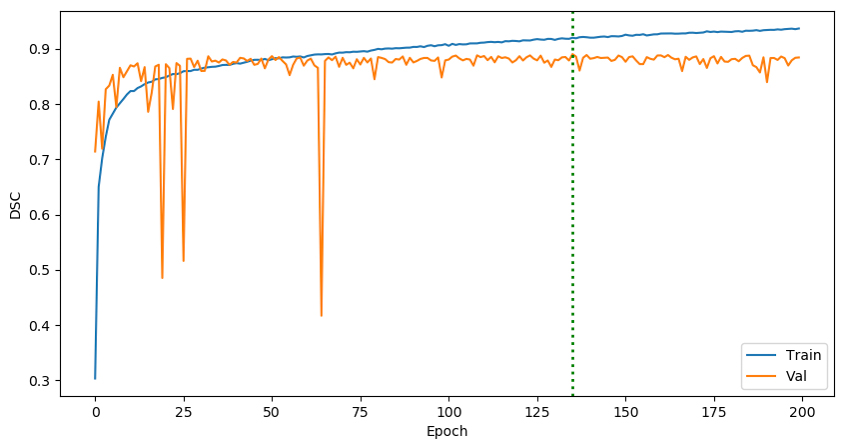}
                \caption{}
                \label{fig:dsc-fold1}
        \end{subfigure}
        \begin{subfigure}{0.41\textwidth}
                \centering
                \includegraphics[width=\textwidth]{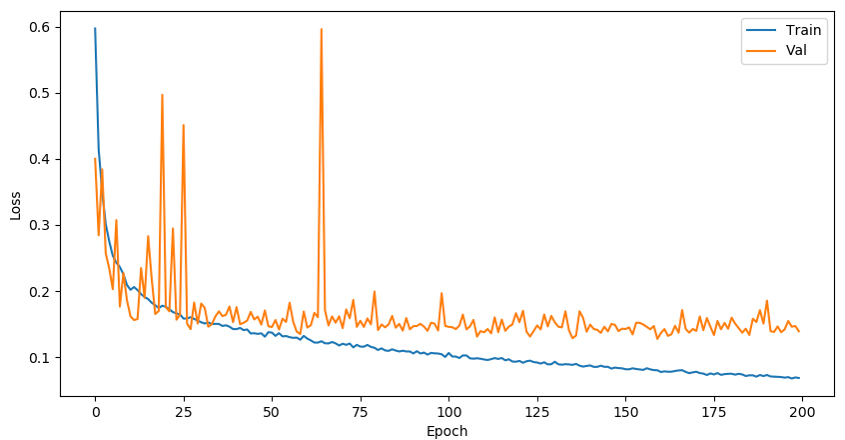}
                \caption{}
                \label{fig:acc-fold1}
        \end{subfigure} %\newline
        
        %%Fold2
        \begin{subfigure}{0.41\textwidth}
                \centering
                \includegraphics[width=\textwidth]{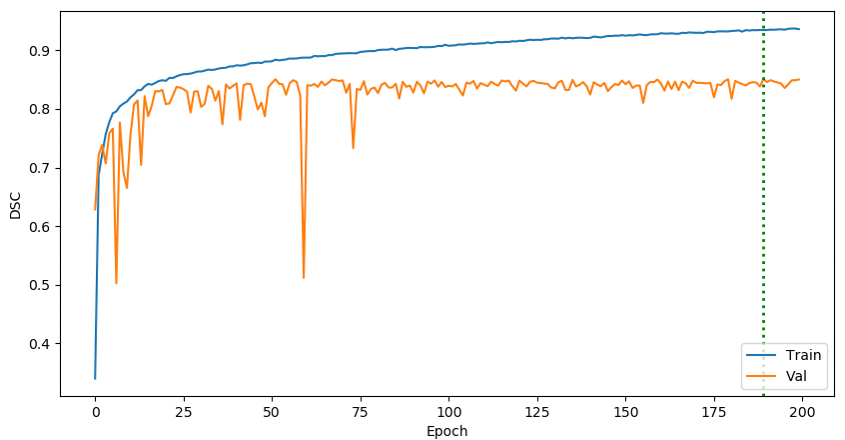}
                \caption{}
                \label{fig:dsc-fold2}
        \end{subfigure}
        \begin{subfigure}{0.41\textwidth}
                \centering
                \includegraphics[width=\textwidth]{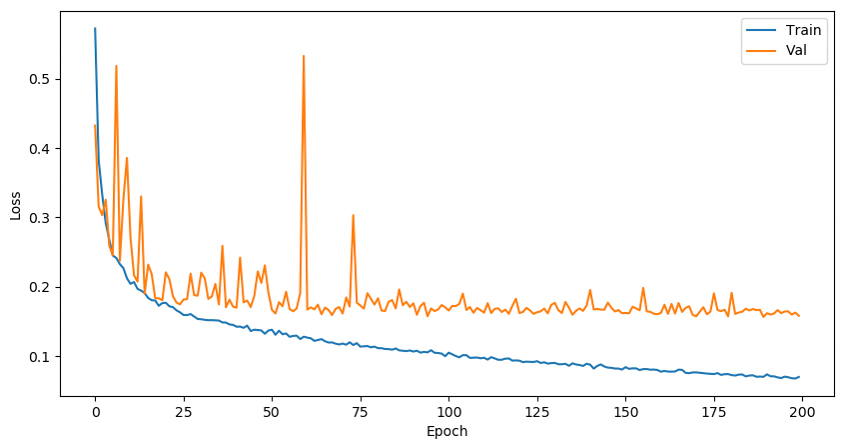}
                \caption{}
                \label{fig:acc-fold2}
        \end{subfigure} %\newline

        %%Fold3
        \begin{subfigure}{0.41\textwidth}
                \centering
                \includegraphics[width=\textwidth]{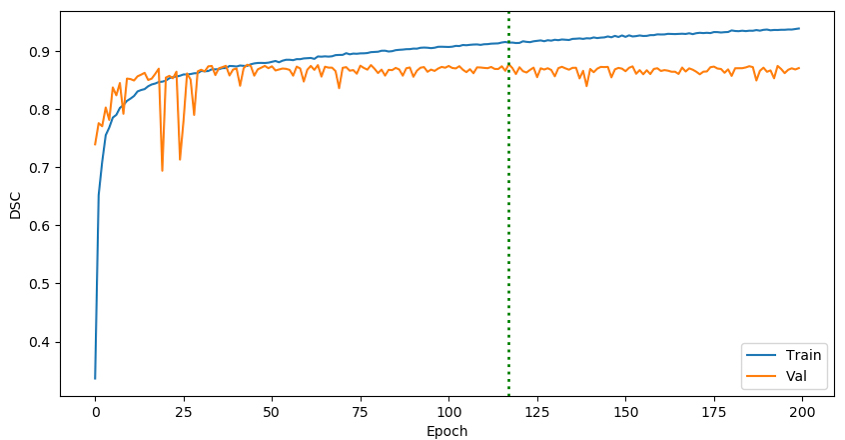}
                \caption{}
                \label{fig:dsc-fold3}
        \end{subfigure}
        \begin{subfigure}{0.41\textwidth}
                \centering
                \includegraphics[width=\textwidth]{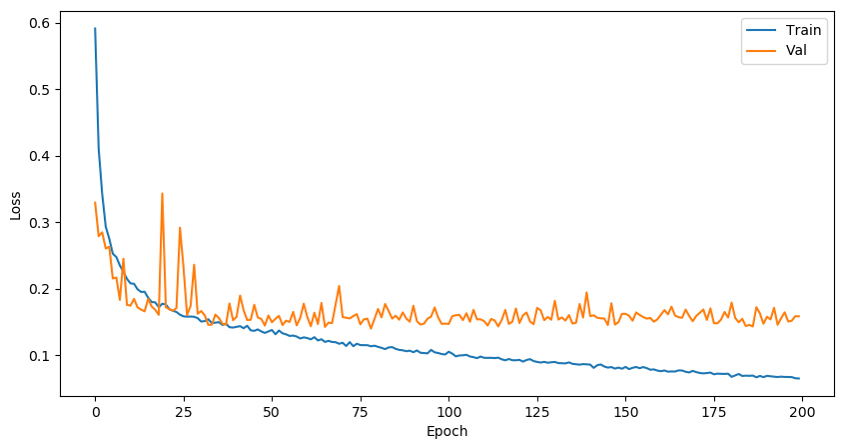}
                \caption{}
                \label{fig:acc-fold3}
        \end{subfigure} %\newline
        
        %%Fold4
        \begin{subfigure}{0.41\textwidth}
                \centering
                \includegraphics[width=\textwidth]{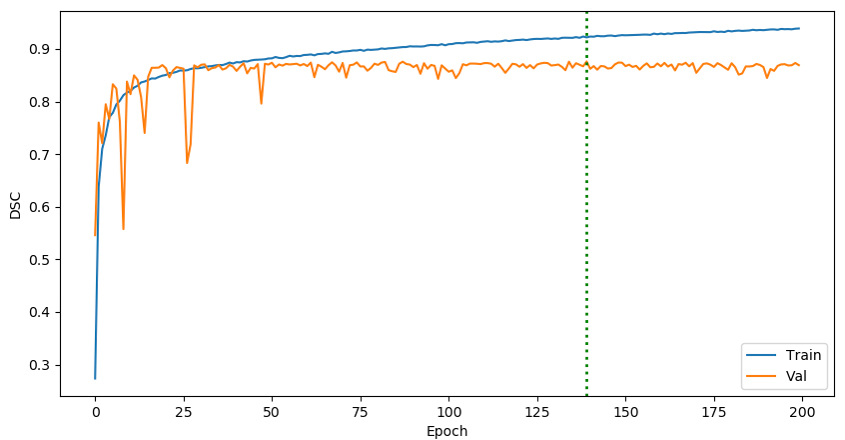}
                \caption{}
                \label{fig:dsc-fold4}
        \end{subfigure}
        \begin{subfigure}{0.41\textwidth}
                \centering
                \includegraphics[width=\textwidth]{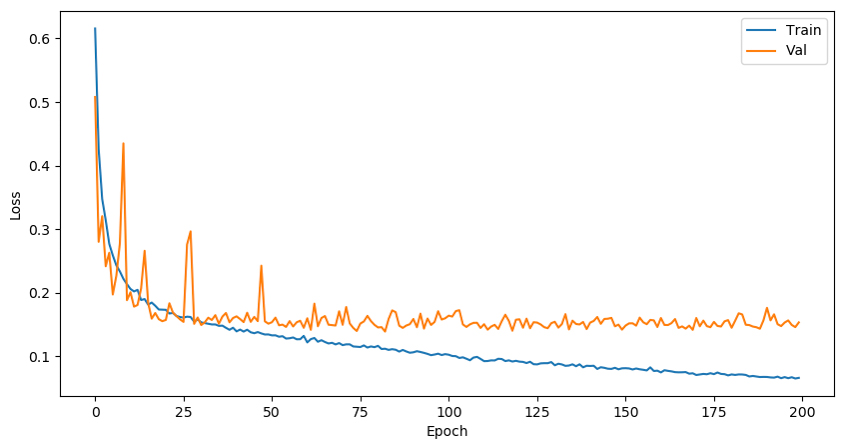}
                \caption{}
                \label{fig:acc-fold4}
        \end{subfigure} %\newline
        
        %%Fold5
        \begin{subfigure}{0.41\textwidth}
                \centering
                \includegraphics[width=\textwidth]{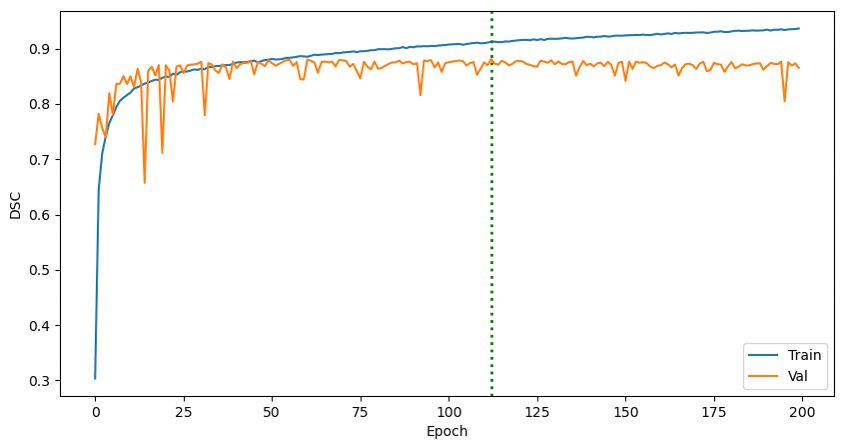}
                \caption{}
                \label{fig:dsc-fold5}
        \end{subfigure}
        \begin{subfigure}{0.41\textwidth}
                \centering
                \includegraphics[width=\textwidth]{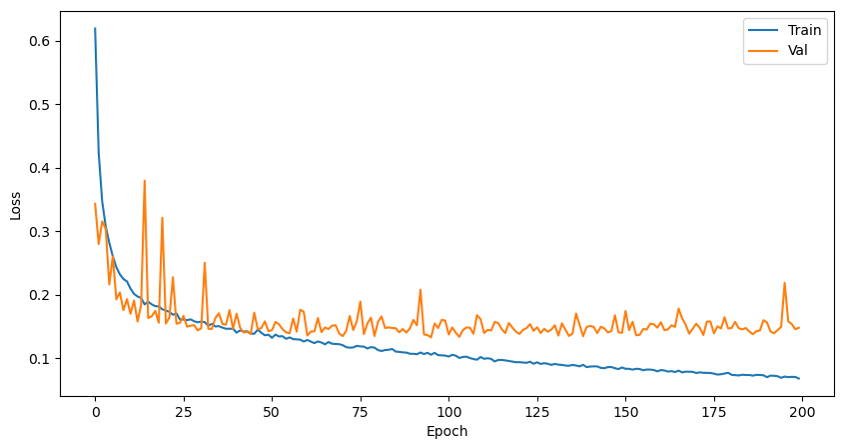}
                \caption{}
                \label{fig:acc-fold5}
        \end{subfigure} %\newline
\caption{Dice (first column) and Loss (second column) produced after $200$ epochs by \textit{Fold1} (a, b), \textit{Fold2} (c, d), \textit{Fold3} (e, f), \textit{Fold4} (g, h), and \textit{Fold5} (i, j). The best model was selected based on the performance of validation set. The green line indicates the peak performance in term of Dice Score on validation set.}
\label{fig:grafici}
\end{figure}

The proposed approach was implemented in Python language (version 3.9.7) using Pytorch \cite{NEURIPS2019_9015} package. All experiments were done on a NVIDIA Quadro RTX 6000 GPU.
The network was evaluated using the Dice loss function, which considers both local and global information. Network training was performed for $200$ epochs well beyond the average converging rate, through the usage of Stochastic Gradient Descent (SGD) \cite{ruder2016overview} as optimizer with an initial learning rate of $1e-4$, a weight decay equal to $1e-4$ and a batch size fixed at $4$. 
%Figure \ref{fig:grafici} shows the graphs of accuracy, loss and Dice obtained from the training considering Fold2 as the configuration.
Figure \ref{fig:grafici} shows the Dice and loss curves obtained during training considering the various folds as configurations. The best model was then selected to perform all tests based on the highest Dice value achieved by the validation set for each fold. 
The training computation time for $200$ epochs was approximately $20$ hours.
No data augmentation was applied during the training process. To demonstrate the absence of overfitting, a subsequent test was performed by applying random transformations to the training data, including flipping and affine transformations of the images. This experiment helped to ensure the model's convergence while avoiding overfitting. The results obtained from the additional training are very similar to those reported in Figure \ref{fig:grafici}. This suggests that the model's performance is robust.

\subsection{General results}
\label{sec:quantitative_results}
To properly evaluate the performances of the proposed approach a set of tests were conducted, in which the employed folds contain multiple combinations of the data.

The averages ± Standard Deviation (SD) of all metrics obtained in the cross-validation test folds are reported in Table \ref{table:table2}. From the comparison of all the results trained in the different folds, it can be concluded that the model achieves the best result in terms of Dice score in \textit{Fold2} ($89$\%). In general, high values are achieved for all metrics and the results being very comparable between all folds.

%%\begin{table}[t]
%%\setlength{\tabcolsep}{3pt}
%%\resizebox{\linewidth}{!}{
%%\begin{tabular}{cccccccc}
\begin{table}[b]

\centering
\resizebox{\linewidth}{!}{
\renewcommand{\arraystretch}{1.1}
\begin{tabular}{c||c||c||c||c||c||c||c||c}

\hline
\hline
\textbf{Fold} & \textbf{Dice} & \multirow{2}{*}{\textbf{Sensitivity}} & \multirow{2}{*}{\textbf{Specificity}} & \multirow{2}{*}{\textbf{IOU}} & \multirow{2}{*}{\textbf{EF}}  & \multirow{2}{*}{\textbf{PPV}} & \multirow{2}{*}{\textbf{NPV}} & \multirow{2}{*}{\textbf{Accuracy}} \\
   
			 \textbf{ID} & \textbf{Score} &  &  &  &  &  & \\
\hline

1       & 0.8448      & 0.8601      & 0.9983      & 0.7314      & 0.0016         & 0.8301      & 0.9987      & 0.9971      \\
2       & \textbf{0.8900}      & 0.8679      & 0.9987      & \textbf{0.8019}      & 0.0012         & \textbf{0.9133}      & 0.9980      & 0.9968      \\
3       & 0.8855      & \textbf{0.9071}      & 0.9995      & 0.7946      & 0.0004         & 0.8650      & \textbf{0.9997}      & \textbf{0.9992}      \\
4       & 0.8101      & 0.7675      & \textbf{0.9997}      & 0.6808      & \textbf{0.0002}         & 0.8577      & 0.9995      & \textbf{0.9992}      \\
5       & 0.8190      & 0.8442      & 0.9992      & 0.6936      & 0.0007         & 0.7953      & 0.9994      & 0.9987      \\

\hline
\textbf{Mean}    & \textbf{0.84±0.03} & \textbf{0.84±0.05} & \textbf{0.99±5e-4} & \textbf{0.74±0.05} & \textbf{8e-4±5e-4}    & \textbf{0.85±0.04} & \textbf{0.99±7e-4} & \textbf{0.99±1e-3} \\ \hline \hline
\end{tabular}
}
\caption{Average of the evaluation metrics for the proposed approach in the different folds for the test data. The last row shows the average among all folds (Average ± SD).}
\label{table:table2}
\end{table}

\begin{figure}[htbp]
        \centering
        %%INIZIO
        \begin{subfigure}[b]{0.23\textwidth}
                \centering
                \includegraphics[width=\textwidth]{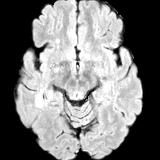}
                \caption{}
                \label{fig:sub-first-slice}
        \end{subfigure}
        \begin{subfigure}[b]{0.23\textwidth}
                \centering
                \includegraphics[width=\textwidth]{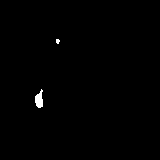}
                \caption{}
                \label{fig:sub-second-gt}
        \end{subfigure}
        \begin{subfigure}[b]{0.23\textwidth}
                \centering
                \includegraphics[width=\textwidth]{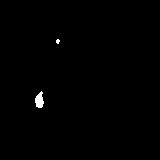}
                \caption{}
                \label{fig:sub-third-pred}
        \end{subfigure}
        \begin{subfigure}[b]{0.23\textwidth}
                \centering
                \includegraphics[width=\textwidth]{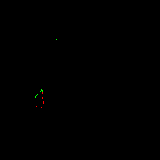}
                \caption{}
                \label{fig:sub-fourth-overlay}
        \end{subfigure} \newline
        
        %%CENTRO

        \begin{subfigure}[b]{0.23\textwidth}
                \centering
                \includegraphics[width=\textwidth]{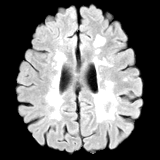}
                \caption{}
                \label{fig:sub2-first-slice}
        \end{subfigure}
        \begin{subfigure}[b]{0.23\textwidth}
                \centering
                \includegraphics[width=\textwidth]{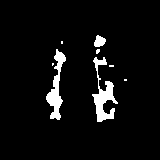}
                \caption{}
                \label{fig:sub2-second-gt}
        \end{subfigure}
        \begin{subfigure}[b]{0.23\textwidth}
                \centering
                \includegraphics[width=\textwidth]{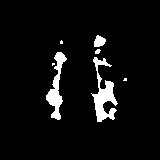}
                \caption{}
                \label{fig:sub2-third-pred}
        \end{subfigure}
        \begin{subfigure}[b]{0.23\textwidth}
                \centering
                \includegraphics[width=\textwidth]{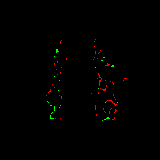}
                \caption{}
                \label{fig:sub2-fourth-overlay}
        \end{subfigure} \newline

        %%FINE

        \begin{subfigure}[b]{0.23\textwidth}
                \centering
                \includegraphics[width=\textwidth]{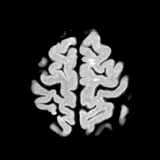}
                \caption{}
                \label{fig:sub3-first-slice}
        \end{subfigure}
        \begin{subfigure}[b]{0.23\textwidth}
                \centering
                \includegraphics[width=\textwidth]{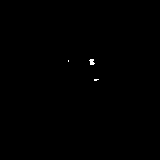}
                \caption{}
                \label{fig:sub3-second-gt}
        \end{subfigure}
        \begin{subfigure}[b]{0.23\textwidth}
                \centering
                \includegraphics[width=\textwidth]{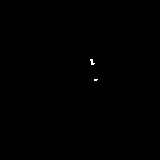}
                \caption{}
                \label{fig:sub3-third-pred}
        \end{subfigure}
        \begin{subfigure}[b]{0.23\textwidth}
                \centering
                \includegraphics[width=\textwidth]{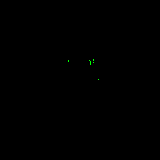}
                \caption{}
                \label{fig:sub3-fourth-overlay}
        \end{subfigure}
        
        \caption{Results of our approach for cases trained with FLAIR images for fold: \textit{Fold2}. The image shows (a), (e), (i) the original slices, (b), (f), (j) the ground truth for each slice, (c), (g), (k) the predictions of our model, (d), (h), (l) false-positive pixels (in red) and false-negative pixels (in green), for slices taken from three different regions of the brain, respectively.} \label{fig:qualitative_results_p2}
\end{figure}

It is possible to appreciate visually the results of the proposed approach,
in the test set, in Figure \ref{fig:qualitative_results_p2}, where is possible to observe the segmentation results on slices of the same patient extracted from three different regions of the brain. Figure \ref{fig:qualitative_results_p2} shows slices of the same subject (a), (e), (i), the ground truth segmentations (b), (f), (j), the segmentations obtained by the proposed approach (c), (g), (k), the false positive and false negative pixels distincted in red and green pixels respectively in (d), (h), (l). 

As can be observed, the predicted lesions mask is very similar to the ground-truth mask, so the proposed approach segments most of the lesions with good accuracy.
False-positive and false-negative pixels mask, confirms how most of the Flair MS lesions were correctly detected by the model.
%\subsection{Comparison test}
%\label{sec:comparison}

The proposed approach has been compared with recent state-of-the-art solutions based on 2D/3D U-Net for MS lesion segmentation. The comparison was done through the mean Dice score between methods on ISBI2015 dataset. Table \ref{table:comparison} shows the results of state-of-the-art, our results achieved in the best fold (\textit{Fold2} in our case) and our mean calculated considering all the involved folds. 

\begin{table}[t]
\centering
\resizebox{\linewidth}{!}{
\renewcommand{\arraystretch}{1.1}
\begin{tabular}{c||c||c||c||c}

\hline
\hline
\textbf{Papers}           & \textbf{Method}                    & \textbf{Dice Score} & \textbf{PPV}  & \textbf{Sensitivity} \\ \hline
Hashemi et al. \cite{hashemi2022delve}  & Attention U-Net           & 0.80       & 0.82 & 0.79        \\
Hashemi et al. \cite{hashemi2022delve}   & U-Net                     & 0.81       & 0.84 & 0.79        \\
Feng et al. \cite{feng2019self}      & 3D U-Net                  & 0.68       & 0.78 & 0.64        \\
Abolvardi et al. \cite{abolvardi2019registration} & 3D U-Net                  & 0.61       & -    & -           \\
Salem et al. \cite{salem2019multiple}     & 2D U-Net                  & 0.64       & 0.79 & 0.57        \\
Aslani et al. \cite{aslani2019multi}    & CNN                       & 0.76       & -    & -           \\
Afzal et al. \cite{afzal2021automatic}     & CNN                       & 0.67       & 0.9  & 0.48        \\
Roy et al. \cite{roy2018multiple}       & CNN                       & 0.56       & 0.6  & -           \\
Zhang et al. \cite{zhang2019multiple}     & FC-DenseNets              & 0.64       & 0.90 & -        \\
Raab et al. \cite{raab2022multimodal}     & 2D U-Net              & 0.77       & -    & -        \\
Kamraoui et al. \cite{kamraoui2022deeplesionbrain}     & 3D U-Net              & 0.67       & 0.84 & -        \\
Sarica et al. \cite{sarica2023dense}     & Dense-Residual U-Net              & 0.66       & 0.86 & -        \\
Our (Mean)       & FC-DenseNet + SA + C-LSTM & 0.84       & 0.85 & 0.84        \\
\textbf{Our (Fold2)}      & \textbf{FC-DenseNet + SA + C-LSTM} & \textbf{0.89}       & \textbf{0.91} & \textbf{0.86}        \\ \hline \hline
\end{tabular}
}
\caption{Mean dice obtained from the proposed approach, compared with Hashemi et al. \cite{hashemi2022delve}, Feng et al. \cite{feng2019self}, Abolvardi et al. \cite{abolvardi2019registration}, Salem et al. \cite{salem2019multiple}, Aslani et al. \cite{aslani2019multi}, Afzal et al. \cite{afzal2021automatic}, Roy et al. \cite{roy2018multiple}, Zhang et al. \cite{zhang2019multiple}, Raab et al. \cite{raab2022multimodal} Kamraoui et al. \cite{kamraoui2022deeplesionbrain}, Sarica et al. \cite{sarica2023dense}. The value in bold indicates the obtained best value; to note that also our mean value overcomes the state-of-the-art. Although some of these methods proposed solutions using all or some of MRI modalities, we report the results of all of them as done by Hashemi et al. \cite{hashemi2022delve} in Table 8.}
\label{table:comparison}
\end{table}

State-of-the-art methods used for sake of comparisons are the following: \cite{kamraoui2022deeplesionbrain,feng2019self,abolvardi2019registration} (3D U-Net architecture), \cite{raab2022multimodal,sarica2023dense,salem2019multiple} (2D U-Net architecture). Also we included \cite{aslani2019multi,afzal2021automatic,roy2018multiple} based on a CNN model while \cite{zhang2019multiple} make use of a Tiramis\'u network by combining slices in the three anatomical planes to capture both global and local contexts. The results of Table \ref{table:comparison} show how our framework improves the results of the state-of-the-art by about $7$ of Dice Score. The results obtained from the method proposed by \cite{hashemi2022delve} are the most similar to ours; \cite{hashemi2022delve} making use of two segmentation networks for MS lesions, an Attention U-Net and a U-Net, and presents the results obtained with both networks on the different MRI acquisition modalities. 

The comparison between our results and \cite{hashemi2022delve} are referred to FLAIR images achieved considering in both the results of the patient corresponding to our \textit{Fold2} as test. Furthermore, it is possible to note how the use of attention mechanisms does not give advantages to \cite{hashemi2022delve}, as Attention U-Net has worse results than simple U-Net. The results obtained with our method in \textit{Fold2} exceed the results obtained by \cite{hashemi2022delve} for both networks.

\subsection{Additional test}
\label{sec:additionaltest}
As previously mentioned, we are in the process of building a new dataset containing FLAIR scans of multiple sclerosis patients with expert-labeled MS lesions.

\begin{figure}[htbp]
        \centering
        %%P1
        \begin{subfigure}[b]{0.23\textwidth}
                \centering
                \includegraphics[width=\textwidth]{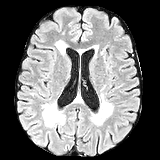}
                \caption{}
                \label{fig:sub-first-slice-a}
        \end{subfigure}
        \begin{subfigure}[b]{0.23\textwidth}
                \centering
                \includegraphics[width=\textwidth]{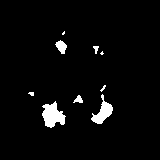}
                \caption{}
                \label{fig:sub-second-gt-a}
        \end{subfigure}
        \begin{subfigure}[b]{0.23\textwidth}
                \centering
                \includegraphics[width=\textwidth]{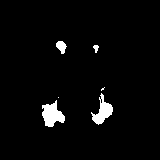}
                \caption{}
                \label{fig:sub-third-pred-a}
        \end{subfigure}
        \begin{subfigure}[b]{0.23\textwidth}
                \centering
                \includegraphics[width=\textwidth]{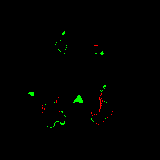}
                \caption{}
                \label{fig:sub-fourth-overlay-a}
        \end{subfigure} \newline
        
        %%P2

        \begin{subfigure}[b]{0.23\textwidth}
                \centering
                \includegraphics[width=\textwidth]{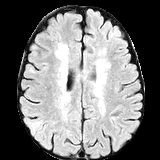}
                \caption{}
                \label{fig:sub2-first-slice-a}
        \end{subfigure}
        \begin{subfigure}[b]{0.23\textwidth}
                \centering
                \includegraphics[width=\textwidth]{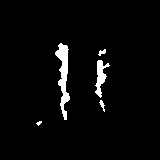}
                \caption{}
                \label{fig:sub2-second-gt-a}
        \end{subfigure}
        \begin{subfigure}[b]{0.23\textwidth}
                \centering
                \includegraphics[width=\textwidth]{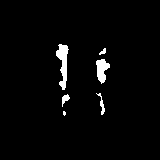}
                \caption{}
                \label{fig:sub2-third-pred-a}
        \end{subfigure}
        \begin{subfigure}[b]{0.23\textwidth}
                \centering
                \includegraphics[width=\textwidth]{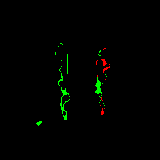}
                \caption{}
                \label{fig:sub2-fourth-overlay-a}
        \end{subfigure} \newline

        %%P3
        \begin{subfigure}[b]{0.23\textwidth}
                \centering
                \includegraphics[width=\textwidth]{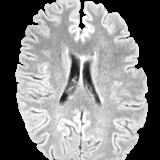}
                \caption{}
                \label{fig:sub3-first-slice-a}
        \end{subfigure}
        \begin{subfigure}[b]{0.23\textwidth}
                \centering
                \includegraphics[width=\textwidth]{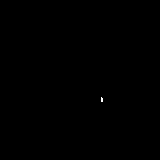}
                \caption{}
                \label{fig:sub3-second-gt-a}
        \end{subfigure}
        \begin{subfigure}[b]{0.23\textwidth}
                \centering
                \includegraphics[width=\textwidth]{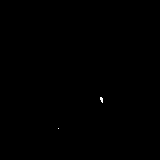}
                \caption{}
                \label{fig:sub3-third-pred-a}
        \end{subfigure}
        \begin{subfigure}[b]{0.23\textwidth}
                \centering
                \includegraphics[width=\textwidth]{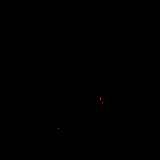}
                \caption{}
                \label{fig:sub3-fourth-overlay-a}
        \end{subfigure} \newline

        \caption{Some of results obtained in different trained networks of our approach tested with three patients of the dataset under construction. The image shows (a), (e), (i) the original slice, (b), (f), (j) the ground-truth for that slice, (c), (g), (k) the prediction of our model, and (d), (h), (l) the false-positive (in red) and false-negative (in green) pixels of P1, P2 and P3, respectively.} \label{fig:qualitative_results_p3test}
\end{figure}

To evaluate the performances of the proposed method (with the model trained on the ISBI-2015 dataset) an additional test was done employing MR FLAIR images from three patients of our in progress dataset. As depicted in Figure \ref{fig:qualitative_results_p3test}, the ground truths of three patients (P1, P2, and P3) were overlaid with the corresponding segmentations obtained by our model and the resulting masks. The high number of false negative pixels indicates that lesion contouring is the task where our network is less accurate. The larger are the lesions the less accurate is the contouring. Table \ref{table:table-additional-test} reports individual fold and average Dice Scores from each patient. Dice Score performances obtained by our method are somehow different when the results from training dataset are compared with the test scans, but the mean Dice Score of 0.7730 achieved for P1 represents a satisfying result in terms of accuracy and lesion segmentation. This discrepancy may be due to differences between acquisition scanners, as a 3.0 T scanner was used for the training images (ISBI-2015), whereas a 1.5 T scanner was employed for testing images. In addition, a reduced performance on test scans may reflect the fact that they were obtained from a single time point, whereas the scans of training set included multiple serial acquisitions for each patients, which improved the accuracy of our automated segmentation method.

\begin{table}[t]
\centering
\tiny
\resizebox{0.9\linewidth}{!}{
\renewcommand{\arraystretch}{1}
\begin{tabular}{c||c||c||c}
\hline
\hline
\textbf{Fold ID} & \textbf{DSC P1}                          & \textbf{DSC P2}                          & \textbf{DSC P3}                         \\ \hline
1       & 0.8023                      & 0.6944                      & 0.5646                     \\
2       & 0.7290                      & 0.6248                      & 0.3443                     \\
3       & 0.8016                      & 0.6097                      & 0.4483                     \\
4       & 0.7954                      & 0.6869                      & 0.5537                     \\
5       & 0.7390                      & 0.5765                      & 0.4021                     \\
\textbf{Mean}    & \textbf{0.7730±0.03} & \textbf{0.6384±0.05} & \textbf{0.4626±0.09} \\ \hline \hline
\end{tabular}
}
\caption{Additional test performed on three different patients extracted from the new dataset, named P1, P2 and P3, respectively, using the best validation model for each Folds.  The results are reported in terms of Dice Score (DSC). The last row shows the average Dice Score across all folds (Average ± SD)}
\label{table:table-additional-test}
\end{table}

\subsection{Ablation studies}
\label{sec:ablation}
In order to explain the reason behind the design of the employed architecture was chosen some ablation studies were done. The proposed network architecture consists of a main backbone to which several modules have been added, then some variants will be presented in this Section.
Specifically, some tests were carried out removing parts of the network or replacing them with others in order to obtain a better explanation of the model behavior and overall achieved performance.
 The purpose is to quantitatively measure the contribution of each parts to the overall model. Starting from a specific model, i.e. the Tiramis\'u network architecture with squeeze and attention layers in the two paths and the unidirectional convolutional LSTM layer in the bottleneck, (shown as FC-DenseNet + SA + C-LSTM in Table  \ref{table:ablation}), three different configurations were considered: Basic Tiramis\'u model (FC-DenseNet in Table \ref{table:ablation}), Tiramis\'u model with the addition of the unidirectional convolutional LSTM level in the bottleneck (FC-DenseNet + C-LSTM in Table \ref{table:ablation}) and the Tiramis\'u model with the addition of the squeeze and attention modules in the two network paths (FC-DenseNet + SA in Table \ref{table:ablation}).
Every configuration was tested on two of the five folds described in section \ref{sec:setup}, chosen on the basis of the results obtained in Section \ref{sec:quantitative_results}. In particular, the two folds with the best and worst Dice Scores in test experiments were chosen, \textit{Fold2} and \textit{Fold4}, respectively.
As can be verified from Table \ref{table:ablation}, the ablation studies demonstrate how SA module always improves the performances while C-LSTM works only if it is in couple with SA. 

\begin{table}[t]
\centering
\resizebox{\linewidth}{!}{
\renewcommand{\arraystretch}{1}
\begin{tabular}{c||c||c||c}
\hline
\hline
\textbf{Architecture}                                    & \textbf{DSC (Fold2)} & \textbf{DSC (Fold4)} & \textbf{DSC Mean} \\ \hline
FC-DenseNet\cite{jegou2017one} & 0.8875      & 0.7989      & 0.8432   \\
FC-DenseNet + C-LSTM                            & 0.8639      & 0.7794      & 0.8216   \\
FC-DenseNet + SA                                & 0.8939      & 0.8048      & 0.8493   \\
FC-DenseNet + SA + C-LSTM                       & 0.8900      & 0.8101      & \textbf{0.8500}   \\ \hline
\hline
\end{tabular}
}
\caption{Ablation studies performed on Folds 2 and 4 with different network configurations employing the ISBI-2015 dataset.}
\label{table:ablation}
\end{table}

%\color{blue} CORREGGI IL VALORE [11], TRASPONI LA TABELLA ED AGGIUNGI IL COMMENTO CHE [11] CHE è QUELLO + SIMILE AL NOSTRO DICE MEDIO RAGGIUNGE QUESTO VALORE CON SOLO 1 PAZIENTE DI TEST, MA NON DA INFO IN MEDIA, MENTRE NOI ABBIAMO UN DICE MEDIO (SU TUTTI I FOLD) CHE COMUNQUE LO BATTE. INOLTRE SCRIVI CHE L'ATTENTION AD [11] NON DA VANTAGGI IN QUANTO LA RETE CON ATTENTION SU FLAIR GLI DA RISULTATI PEGGIORI MENTRO NOI (CHE SAPPIAMO COME USARE L'ATTENTION) ABBIAMO UN BOOST DEI RISULTATI\color{black}

%To demonstrate the effectiveness of the proposed solution, the results obtained from our framework was compared with recent state-of-the-art solutions using U-Net, both 2D and 3D, for MS lesion segmentation. In order to fairly compare the results obtained, we focused only on the work using the ISBI2015 dataset. We compare the results obtained in terms of mean Dice score. Specifically, Table \ref{table:comparison} shows the results achieved for the fold that obtained the highest value, namely Fold2. The effectiveness of the proposed model is compared with \cite{feng2019self}, \cite{abolvardi2019registration} who propose a 3D U-Net architecture, while the authors in \cite{salem2019multiple} use a 2D U-Net architecture, \cite{aslani2019multi}, \cite{afzal2021automatic}, \cite{roy2018multiple} propose CNN model, \cite{zhang2019multiple} propose a Tiramisù network by combining slices in the three anatomical planes to capture both global and local contexts.

\section{Conclusion and future work}
\label{sec:conclusion}

In this paper we proposed a new framework to address the problem of MS lesion segmentation on MRI in the effort to facilitate the estimation of disease burden overtime. In particular, our approach is based on an extension of the U-net neural network. The proposed method demonstrated to be more accurate than state-of-the-art methods, boosting results by exploiting a dedicated attention mechanism. It is worth noting that the simple insertion of attention does not always improves results \cite{hashemi2022delve}, whilst only a dedicated solution, as the novel one presented in this paper, could be able to provide substantial improvement. The effectiveness and robustness of the technique was demonstrated for the first time on patients never employed for the training of the model. The high level of Dice Score, obtained by the proposed method on this particular sample, is of utter importance in demonstrating the generalizing capabilities of the solution, as it is not dependent to a specific acquisition hardware and method. To further investigate these capabilities, we are continuing the acquisition campaign with the aim to have new samples and enrich the comparison dataset. \\
Furthermore, the lesion segmentation framework proposed in the paper uses recent AI methodologies to estimate the level of progression of the MS disease by recognizing automatically the lesions in MRI images. The obtained data about the quantitative MRI lesion load will be used in the UISS framework, which has the aim to model and simulate the progression of MS lesions as well as to predict the immune response to specific treatments. A potential future direction for the research is to explore the possibility of replacing the recurrent layers with 3D convolutions to enhance the performance of the network.

%\ref{sec:sample:appendix}.

%\appendix

\section*{Acknowledgements}
\label{sec:acknowledgements}
Alessia Rondinella is a PhD candidate enrolled in the National PhD in Artificial Intelligence, XXXVII cycle, course on Health and life sciences, organized by Università Campus Bio-Medico di Roma.

Authors Francesco Guarnera, Giulia Russo, and Francesco Pappalardo were supported by the European Commission through the H2020 project “In Silico World: Lowering barriers to ubiquitous adoption of In Silico Trials” (topic SC1-DTH-06-2020, grant ID 101016503).

Experiments were carried out thanks to the hardware and software granted and managed by iCTLab S.r.l. - Spinoff of University of Catania.

 \bibliographystyle{elsarticle-num} 
 \bibliography{cas-refs}





% \end{thebibliography}
\end{document}